\documentclass[aps,prb,amsmath,superscriptaddress,twocolumn,longbibliography]{revtex4-2}

\usepackage{graphicx}
\usepackage{amsmath}
\usepackage{amsfonts}
\usepackage{amssymb}
\usepackage{braket}
\usepackage{bm}
\usepackage{multirow}
\usepackage{color}
\usepackage[normalem]{ulem}
\usepackage{mathrsfs}
\usepackage{mathtools}
\usepackage{float}

\makeatletter

\newcommand{\Rmnum}[1]{\expandafter\@slowromancap\romannumeral #1@}
\makeatother

\usepackage[breaklinks]{hyperref}
\hypersetup{colorlinks=true, linkcolor=blue, citecolor=blue, filecolor=blue, urlcolor=blue}

\AtBeginDocument{%
    \newwrite\bibnotes
    \def\bibnotesext{Notes.bib}
    \immediate\openout\bibnotes=\jobname\bibnotesext
    \immediate\write\bibnotes{@CONTROL{REVTEX41Control}}
    \immediate\write\bibnotes{@CONTROL{%
    apsrev41Control,author="08",editor="1",pages="1",title="0",year="1"}}
     \if@filesw
     \immediate\write\@auxout{\string\citation{apsrev41Control}}%
    \fi
}%

\begin{document}

\title{Distinction Between Transport and R\'enyi Entropy Growth in Kinetically Constrained Models}

\author{Zhi-Cheng Yang}
\email{zcyang19@pku.edu.cn}
\affiliation{School of Physics, Peking University, Beijing 100871, China}
\affiliation{Center for High Energy Physics, Peking University, Beijing 100871, China}

\date{\today}

\begin{abstract}

Conservation laws and the associated hydrodynamic modes have important consequences on the growth of higher R\'enyi entropies in isolated quantum systems. It has been shown in various random unitary circuits and Hamiltonian systems that the dynamics of the R\'enyi entropies in the presence of a U(1) symmetry obey $S^{(n\geq 2)}(t) \propto t^{1/z}$, where $z$ is identified as the dynamical exponent characterizing transport of the conserved charges. Here, however, we demonstrate that this simple identification may not hold in certain quantum systems with kinetic constraints. In particular, we study two types of U(1)-symmetric quantum automaton circuits with XNOR and Fredkin constraints, respectively. We find numerically that while spin transport in both models is subdiffusive, the second R\'enyi entropy grows diffusively in the XNOR model, and superdiffusively in the Fredkin model.
For systems with XNOR constraint, this distinction arises since the spin correlation function can be attributed to an emergent tracer dynamics of tagged particles, whereas the R\'enyi entropies are constrained by collective transport of the particles. Our results suggest that care must be taken when relating transport and entanglement entropy dynamics in generic quantum systems with conservation laws.

\end{abstract}

\maketitle

{\it Introduction.---}  Fundamental questions on non-equilibrium dynamics and thermalization in isolated quantum systems have gained revived interests over the past decade~\cite{rigol2008thermalization, d2016quantum, gogolin2016equilibration}, thanks in part to the advances in quantum technology, lending people with novel experimental platforms with unprecedented level of controllability~\cite{preskill2018quantum}. In thermalizing systems with one or a few conserved quantities (e.g. energy, charge, magnetization), an effective hydrodynamic description emerges, which captures transport of the conserved charges beyond local equilibrium timescales~\cite{PhysRevB.73.035113, PhysRevA.89.053608}. The emergence of hydrodynamics (and its generalized version) in quantum systems has recently been widely explored in the context of operator spreading~\cite{PhysRevX.8.031057, PhysRevX.8.031058}, models with unusual fractonic constraints~\cite{PhysRevResearch.2.033124, glorioso2022breakdown}, and integrable systems in one dimension~\cite{PhysRevLett.121.230602,ljubotina2017spin, gopalakrishnan2019anomalous, PhysRevLett.127.057201, PhysRevX.11.031023, PhysRevLett.122.127202, de2021subdiffusive}.

Recently, an intriguing connection between hydrodynamic modes and quantum information dynamics was unveiled. In particular, it was shown in several U(1)-symmetric random unitary circuit models and Hamiltonian systems with energy conservation that while the von Neumann entropy grows ballistically~\cite{PhysRevLett.111.127205}, higher R\'enyi entropies $S^{(n\geq 2)}$ can only grow as fast as $\sqrt{t}$ due to the existence of slow diffusive modes~\cite{PhysRevLett.122.250602, vznidarivc2020entanglement, huang2020dynamics, PhysRevResearch.2.033020}. This connection was later generalized to transport types other than diffusion in a setup involving U(1)-symmetric long-range Clifford circuits~\cite{richter2022transport}. By varying the power $\alpha$ controlling the probability $P(r) \propto r^{-\alpha}$ of gates spanning a distance $r$, transport in this system can be diffusive, superdiffusive, ballistic, or superballistic. The R\'enyi entropy growth is found to be always consistent with 
\begin{equation}
S^{(n\geq 2)} (t) \propto t^{1/z}, 
\label{eq:renyi_z}
\end{equation}
where $z$ is the $\alpha$-dependent dynamical exponent characterizing transport of the conserved charge.

\begin{figure}[!t]
\includegraphics[width=.5\textwidth]{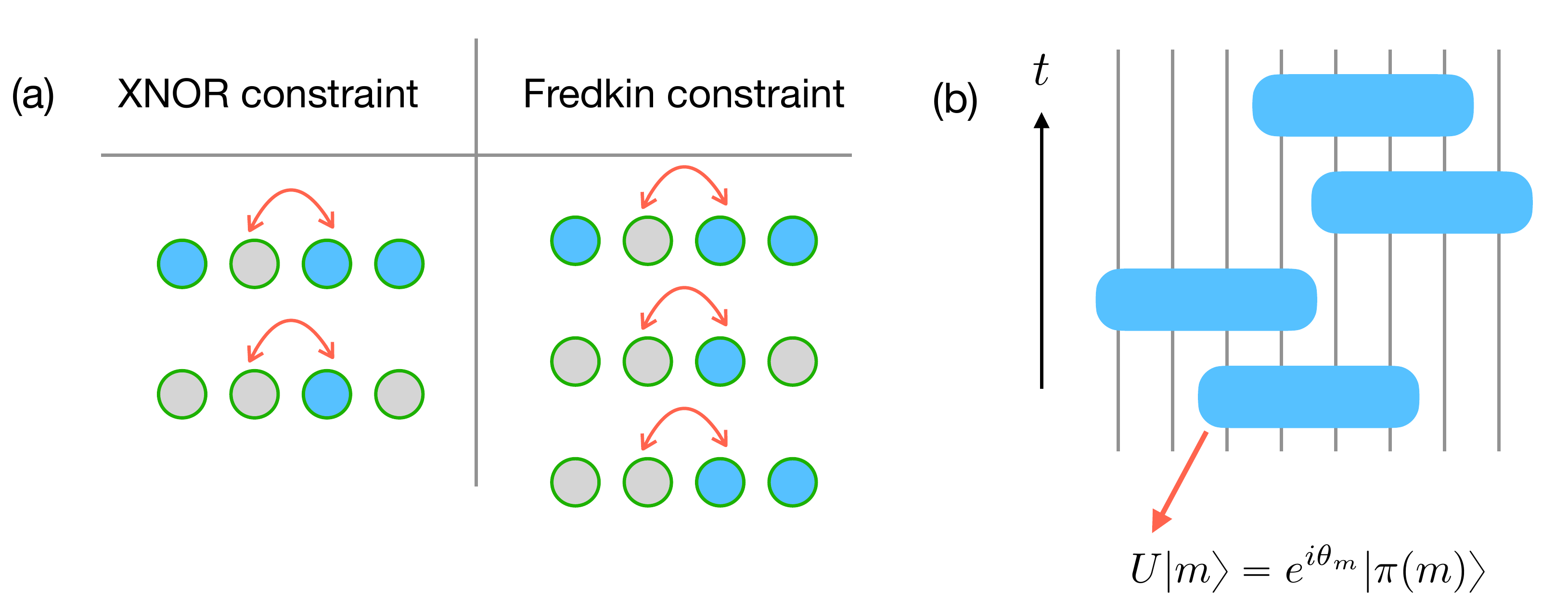}
\caption{(a) Allowed dynamical moves for the XNOR and Fredkin constraint, respectively. A blue dot denotes an occupied site, and a grey dot denotes an unoccupied site. (b) Setup for the QA circuit. At each step, a four-site QA gate is applied on randomly chosen four consecutive sites. For a system of $L$ qubits, one time step consists of $L$ gates.}
\label{fig:setup} 
\end{figure}

The above results point towards an appealing scenario that higher R\'enyi entropy growth and transport in generic quantum systems with conservation laws can always be related via Eq.~(\ref{eq:renyi_z}). In this work, we show that relation~(\ref{eq:renyi_z}) in fact may not hold in certain quantum systems with kinetic constraints. We study two types of U(1)-symmetric quantum automaton (QA) circuits with XNOR and Fredkin constraints, respectively (Fig.~\ref{fig:setup}). Both types of constraints lead to subdiffusive transport~\cite{PhysRevLett.127.230602}, and XNOR constraint further gives rise to Hilbert-space fragmentation~\cite{PhysRevLett.124.207602, PhysRevB.103.L220304}. We show that both the dynamical spin correlation function and the second R\'enyi entropy $S^{(2)}$ can be efficiently calculated numerically in QA circuits. Subdiffusive transport in both models is demonstrated via calculation of the spin correlators, yielding $z=4$ for the XNOR QA circuits and $z \simeq \frac{8}{3}$ for the Fredkin QA circuit, consistent with previous results on Haar random circuits~\cite{PhysRevLett.127.230602}. On the other hand, we find that the second R\'enyi entropy grows diffusively $S^{(2)}(t) \propto t^{1/2}$ for the XNOR circuit and superdiffusively $S^{(2)}(t) \propto t^{0.6}$ for the Fredkin circuit. We argue for the XNOR circuit that this distinction arises since the spin correlation function can be attributed to an emergent tracer dynamics of tagged particles~\cite{feldmeier2022emergent}, whereas the R\'enyi entropies are instead constrained by collective transport of the particles. The scenario is also reminiscent of the distinction between spin and energy transport in the integrable easy-axis $XXZ$ model: while spin transport is diffusive due to screening of the quasiparticles (magnons), energy transport is still ballistic since quasiparticles do propagate ballistically~\cite{ljubotina2017spin, PhysRevLett.106.220601, PhysRevLett.122.127202}. Our results thus provide a concrete example suggesting that care must be taken when relating transport and entanglement entropy dynamics using Eq.~(\ref{eq:renyi_z}) in generic quantum systems with conservation laws.

{\it Quantum automaton circuits and kinetic constraints.--} A QA gate or circuit can be defined via its action on a computational basis state $|m\rangle$~\cite{PhysRevB.100.214301, PRXQuantum.2.010329, PhysRevB.102.224311, gopalakrishnan2018facilitated, PhysRevB.105.064306, han2022entanglement}: $U|m\rangle = e^{i\theta_m} |\pi(m)\rangle$, where $\pi(m)$ is an element of the permutation group acting on the computational basis states, and $\theta_m$ is a random phase depending on the particular state. Apparently, a QA circuit consists of a restricted subset of unitaries and cannot generate entanglement when applied to a product state in the computational basis. However, when acting on an initial state with all qubits initialized in the $|+\hat{x}\rangle$ direction:
\begin{equation}
|\psi_0\rangle = \bigotimes_{i=1}^L |+\hat{x}\rangle_i = \frac{1}{\sqrt{2^L}} \sum_m |m\rangle,
\label{eq:initial}
\end{equation}
a QA circuit can produce highly-entangled states by adding random phases to each computational basis state: $U|\psi_0\rangle = 2^{-L/2}\sum_m e^{i\theta_m}|m\rangle$. The choice of initial state~(\ref{eq:initial}) followed by unitary evolutions generated by QA circuits makes the numerical simulation of various quantities of physical interests tractable, as we will see below.

Kinetic constraints on quantum dynamics are encoded in each elementary gate, and more specifically, in the permutation group element $\pi(m)$. The allowed dynamical moves in the XNOR and Fredkin circuits are summarized in Fig.~\ref{fig:setup}(a). Both models have a U(1) symmetry associated with the total particle number or magnetization.
The XNOR constraint allows hopping of a particle between sites $i$ and $i+1$ only if the two further-neighboring sites $i-1$ and $i+2$ are both occupied or empty. Such a constraint further keeps the total number of domian walls conserved, and gives rise to a fragmented Hilbert space with exponentially many disconnected subsectors~\cite{PhysRevLett.124.207602}. The Fredkin constraint allows hopping if either site $i+2$ is occupied or site $i-1$ is empty. The resulting dynamical moves are analogous to those generated via a Fredkin gate~\cite{chen2017gapless, PhysRevLett.127.230602}, hence the name. These local kinetic constraints in our circuit models are implemented using four-qubit QA gates, where the permutation group elements $\pi(m)$ are chosen to respect the constraints. At each step, a QA gate with randomly drawn phases is applied on randomly chosen four consecutive sites, as shown in Fig.~\ref{fig:setup}(b). For a system of $L$ qubits, one unit time step consists of $L$ gates.

{\it Spin correlation functions.--} Transport properties of the conserved charge can be probed via the infinite-temperature dynamical spin correlation function: $C(x,t) := \langle \hat{S}^z_x(t) \hat{S}^z_0(0)\rangle=2^{-L}{\rm Tr}[\hat{S}^z_x(t) \hat{S}^z_0(0)]$. Ref.~\cite{PhysRevLett.127.230602} considers Haar random local unitaries and maps the calculation of $C(x,t)$ to a classical stochastic process under Haar averaging. We show that the deterministic nature of our QA circuit evolution further simplifies this connection. Indeed, it is straightforward to see that
\begin{eqnarray}
C(x,t) &=& \frac{1}{2^L} \sum_n \langle n | \hat{U}^\dagger \hat{S}^z_x \hat{U} \hat{S}^z_0 |n\rangle  \nonumber \\
&=& \frac{1}{2^L} \sum_n S^z_{n,0} S^z_{\pi(n),x}  \nonumber  \\
&=& \langle \psi_0| \hat{S}^z_x(t) \hat{S}^z_0(0) |\psi_0\rangle,
\label{eq:correlation}
\end{eqnarray}
where $S^z_{n,x}$ denotes the $z$-component spin value of the $x$-th site in configuration $|n\rangle$. Notice that $C(x,t)$ for our QA circuit coincides with the correlator evaluated under the initial state~(\ref{eq:initial}). Therefore, one can efficiently evaluate the spin correlators via sampling from the classical stochastic process with the same kinetic constraint according to Eq.~(\ref{eq:correlation}).

\begin{figure}[!t]
\includegraphics[width=.5\textwidth]{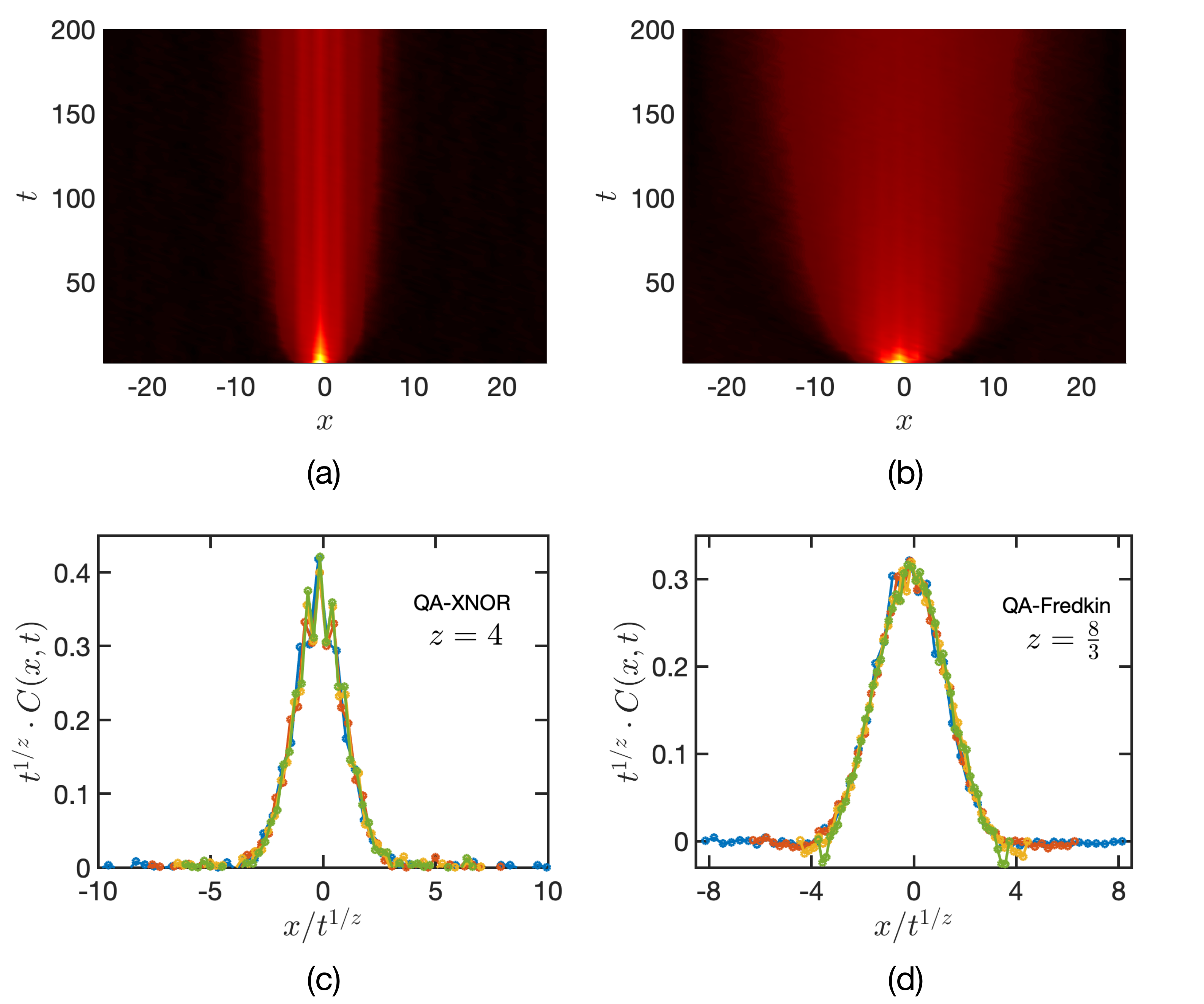}
\caption{Top panel: spatial-temporal profile of $C(x,t)$ for (a) the XNOR model and (b) the Fredkin model. Bottom panel: data collapse of $t^{1/z}\cdot C(x,t)$ versus $x/t^{1/z}$ yields (c) $z=4$ for the XNOR model and (d) $z\simeq \frac{8}{3}$ for the Fredkin model. Numerical simulations are performed for systems with $L=50$ qubits and averaged over $10^4$ circuit realizations.}
\label{fig:correlation} 
\end{figure}

In Figs.~\ref{fig:correlation}(a)\&(b), we show the spatial-temporal profile of $C(x,t)$ for both models. Since the spin correlators should obey the scaling form $C(x,t) = t^{-1/z} f(x/t^{1/z})$, we attempt data collapses by plotting $t^{1/z}\cdot C(x,t)$ versus $x/t^{1/z}$, which allows us to extract the dynamical exponents. Data collapses shown in Figs.~\ref{fig:correlation}(c)\&(d) are indeed consistent with the scaling form, yielding $z=4$ for the XNOR model, and $z\simeq \frac{8}{3}$ for the Fredkin model, demonstrating subdiffusive transport in both models.

{\it Tracer dynamics.--} While a clear understanding of the universality class $z\simeq \frac{8}{3}$ in the Fredkin model remains unknown at the moment, the origion of $z=4$ in the XNOR model can be explained in terms of an emergent tracer dynamics of tagged particles. Let us first give a heuristic argument for the observed subdiffusion. As a result of the constraint, the only mobile objects in the XNOR model are magnons, whose total number is conserved. Consider a single mobile magnon moving in the background of a fixed spin configuration. This magnon propagates diffusively through a domain of up spins as a minority down spin, and then through a domain of down spins as a minority up spin. On average, motion of the magnon carries zero current and transports zero net charge. Therefore, diffusive contribution to charge transport is zero, 
and the leading-order contribution comes from fluctuations of the magnetization over the distance traveled by the magnon, which is indeed subdiffusive with $z=4$~\cite{de2021subdiffusive}.

\begin{figure}[!t]
\includegraphics[width=.3\textwidth]{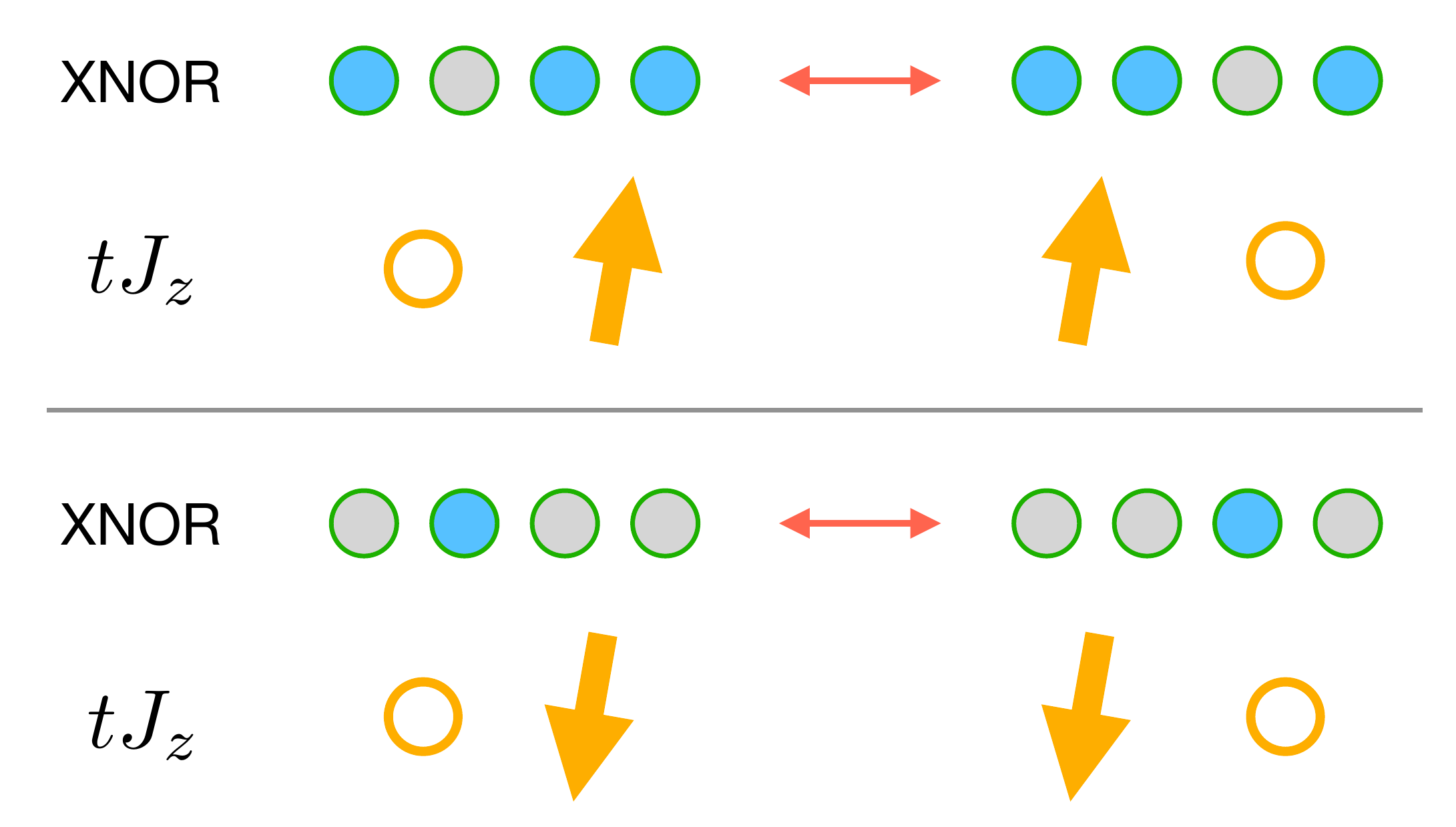}
\caption{Mapping the XNOR model to the $tJ_z$ model. One can define a spin operator $\hat{\Sigma}^z_x$ of the $tJ_z$ model from two neighboring sites in the XNOR model: $\hat{\Sigma}^z_x := \frac{1}{2}(\hat{S}^z_x + \hat{S}^z_{x+1})$. $\Sigma^z=0$ indicates a hole in the $tJ_z$ model.}
\label{fig:tJz} 
\end{figure}

We proceed with a more precise explanation of $z=4$ by mapping to a $tJ_z$ model~\cite{feldmeier2022emergent, PhysRevB.101.125126}. The mapping is illustrated in Fig.~\ref{fig:tJz} (see Ref.~\cite{feldmeier2022emergent} for a more detailed discussion on the mapping, and in particular, mapping of the spin correlators). Like the $tJ$ model, the $tJ_z$ model consists of spinful particles hopping on a lattice, but with the Heisenberg term replaced by a coupling that only involves the $S^z$ component. Therefore, the spin pattern of the particles remains unchanged, which imposes a nearest-neighbor hardcore exclusion constraint for the particles. One can define a spin operator $\hat{\Sigma}^z_x$ of the $tJ_z$ model from two neighboring sites in the XNOR model: $\hat{\Sigma}^z_x := \frac{1}{2}(\hat{S}^z_x + \hat{S}^z_{x+1})$. It is understood that $\Sigma^z=0$ indicates a hole in the $tJ_z$ model. One can readily check that the allowed dynamical moves in the XNOR model directly translates into particles hopping in the $tJ_z$ model, subject to a nearest-neighbor exclusion constraint such that the spin pattern remains unchanged. We thus have a QA version of the $tJ_z$ model, which is equivalent to the original XNOR model. An alternative way of interpreting the connection between the XNOR and the $tJ_z$ model is in terms of the ``root configurations" discussed in Ref.~\cite{PhysRevLett.124.207602}, where a representative state in each connected subsector can be constructed by appending a $k$-magnon state to a frozen configuration. In this language, the dynamics can be described as hopping of magnons (either as an up spin or a down spin) in the background of a frozen configuration.

Let us now argue that the spin correlator $C(x,t)$ in the original XNOR model is described by an emergent tracer dynamics of a tagged particle using the spin correlator in the $tJ_z$ QA model. A state in the $tJ_z$ model can be labeled by $|{\bm r}, {\bm \sigma}\rangle$, where ${\bm r}$ denotes the locations of the particles, and ${\bm \sigma}$ their spins. Notice that $\hat{\Sigma}_x^z = \sum_{i=1}^{N_p} \delta_{{\bm r}_i, x} \hat{\sigma}_i$, with $N_p$ being the total number of particles, we have
\begin{eqnarray}
&&\frac{1}{\mathcal{N}} {\rm Tr} \left[ \hat{\Sigma}^z_x(t) \hat{\Sigma}^z_0(0) \right] = \frac{1}{\mathcal{N}} \sum_{{\bm r}, {\bm \sigma}} \langle {\bm r}, {\bm \sigma}| \hat{U}^\dagger \hat{\Sigma}^z_x \hat{U} \hat{\Sigma}^z_0 |{\bm r}, {\bm \sigma}\rangle  \nonumber \\
&=& \frac{1}{\mathcal{N}} \sum_{{\bm r}, {\bm \sigma}} \sum_{i,j} \delta_{{\bm r}_i, 0} \delta_{\pi({\bm r})_j, x} \sigma_i \sigma_j   \nonumber  \\
&=& \frac{1}{\mathcal{N}_p} \sum_{{\bm r}} \sum_{i,j} \delta_{{\bm r}_i, 0} \delta_{\pi({\bm r})_j, x} \frac{1}{\mathcal{N}_s}\sum_{{\bm \sigma}} \sigma_i \sigma_j   \nonumber  \\
&=&   \frac{1}{\mathcal{N}_p} \sum_{{\bm r}} \sum_{i} \delta_{{\bm r}_i, 0} \delta_{\pi({\bm r})_i, x},
\label{eq:tracer}
\end{eqnarray}
where the normalization factor $\mathcal{N} \equiv \mathcal{N}_p \mathcal{N}_s$, with $\mathcal{N}_s = 2^{N_p}$ and $\mathcal{N}_p = {L \choose N_p}$, and we have used the fact that $\mathcal{N}_s^{-1}\sum_{\bm \sigma} \sigma_i \sigma_j = \delta_{ij}$. It is clear from Eq.~(\ref{eq:tracer}) that the spin correlator receives contribution from trajectories where the $i$-th particle starting at the origin initially ends up at position $x$ at time $t$. Upon circuit averaging, this becomes precisely the probability distribution of the motion of a tagged particle. In the nearest-neighbor simple exclusion process considered here, motion of a tagged particle in an environment with a finite density of other particles is known to be subdiffusive with $z=4$~\cite{PhysRevB.18.2011, PhysRevB.28.5711}. Thus, we expect that the original spin correlation function $C(x,t)$ in the XNOR model will also exhibit the same dynamical exponent.

\begin{figure}[!t]
\includegraphics[width=.45\textwidth]{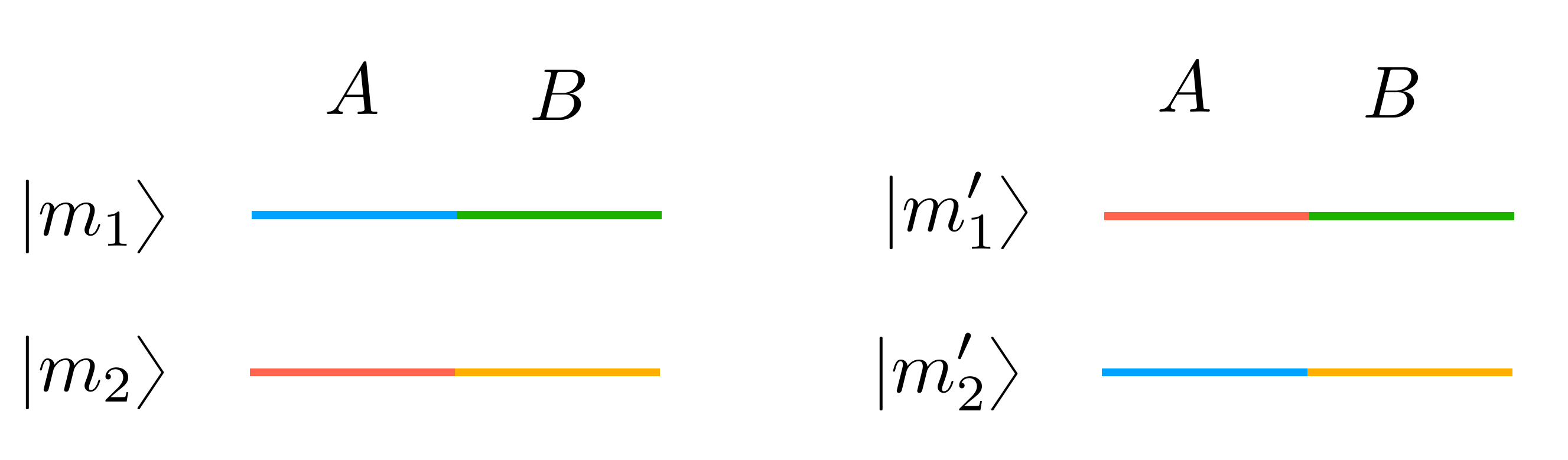}
\caption{Illustration of the two pairs of configurations involved in the calculation of the purity Eq.~(\ref{eq:purity}). $|m_1'\rangle$ and $|m_2'\rangle$ are obtained from $|m_1\rangle$ and $|m_2\rangle$ by exchanging their bitstring configurations in subsystem $A$.}
\label{fig:purity} 
\end{figure}

{\it Second R\'enyi entropy.--}We next turn to the higher R\'enyi entropies $S^{(n)} = [1/(1-n)]{\rm ln} \ {\rm tr}(\rho_A^n)$, with $\rho_A = {\rm tr}_B |\psi\rangle \langle \psi|$ being the reduced density matrix of subsystem $A$ under a bipartitioning of the entire system. We will focus on the second R\'enyi entropy $n=2$ in what follows. To compute $S^{(2)}$, notice that the purity can be written as~\cite{PhysRevB.102.224311, PhysRevB.105.064306, han2022entanglement}
\begin{equation}
{\rm tr}(\rho_A^2) = \langle \psi_0|_1 \langle \psi_0|_2 (U^\dagger \otimes U^\dagger)\ {\rm SWAP}_A\ (U\otimes U)|\psi_0\rangle_1 |\psi_0\rangle_2,   
\end{equation}
where ${\rm SWAP}_A$ exchanges the two replicas within subsystem $A$: ${\rm SWAP}_A |m_A m_B\rangle_1 |n_A n_B\rangle_2 = |n_A m_B\rangle_1 |m_A n_B\rangle_2$. Inserting a resolution of the identity and using properties of the QA circuit, we have
\begin{eqnarray}
{\rm tr}(\rho_A^2) &=& \sum_{m_1, m_2}  \langle \psi_0|_1 \langle \psi_0|_2 (U^\dagger \otimes U^\dagger)\ {\rm SWAP}\ |m_1\rangle |m_2\rangle  \nonumber \\
 && \langle m_1 |\langle m_2| \ (U\otimes U)|\psi_0\rangle_1 |\psi_0\rangle_2  \nonumber \\
&=&  \sum_{m_1, m_2}  \langle \psi_0|_1 \langle \psi_0|_2 (U^\dagger \otimes U^\dagger)\ |m'_1\rangle |m'_2\rangle  \nonumber \\
&& \langle m_1 |\langle m_2| \ (U\otimes U)|\psi_0\rangle_1 |\psi_0\rangle_2   \nonumber \\
&=& \frac{1}{4^L} \sum_{m_1, m_2} e^{i \Theta_{1,m}} e^{i \Theta_{2,m}}.
\label{eq:purity}
\end{eqnarray}
In the above equations, $|m_1'\rangle$ and $|m_2'\rangle$ are obtained from $|m_1\rangle$ and $|m_2\rangle$ by exchanging their bitstring configurations within subsystem $A$, as illustrated in Fig.~\ref{fig:purity}. The automaton property of the circuit then indicates that the forward and backward unitary evolutions on the replicated Hilbert space $U^{\dagger \otimes 2} |m_1'\rangle |m_2'\rangle$ and $\langle m_1|\langle m_2| U^{\otimes 2}$ simply accumulate phases $\Theta_{1,m}$ and $\Theta_{2,m}$, which explains Eq.~(\ref{eq:purity}). To gain some further intuitions behind Eq.~(\ref{eq:purity}), notice that for most pairs of configurations, $|m_1'\rangle \neq |m_1\rangle$ and $|m_2'\rangle \neq |m_2\rangle$ after the SWAP. For evolutions that are long enough, the phases accumulated from the forward and backward evolutions between these pairs are randomized and will in general cancel out. In the long time limit, only pairs of configurations with identical bitstrings within subsystem $A$ lead to the same pair after the SWAP, and hence contribute to the purity. Counting the total number of such pairs gives ${\rm min}[{\rm tr}(\rho_A^2)] = \frac{2^{L_A} 2^{2L_B}}{4^L} =\frac{1}{2^{L_A}}$, and hence ${\rm max}[S^{(2)}]= L_A {\rm ln}2$, consistent with that of a maximally entangled state.

\begin{figure}[!t]
\includegraphics[width=.5\textwidth]{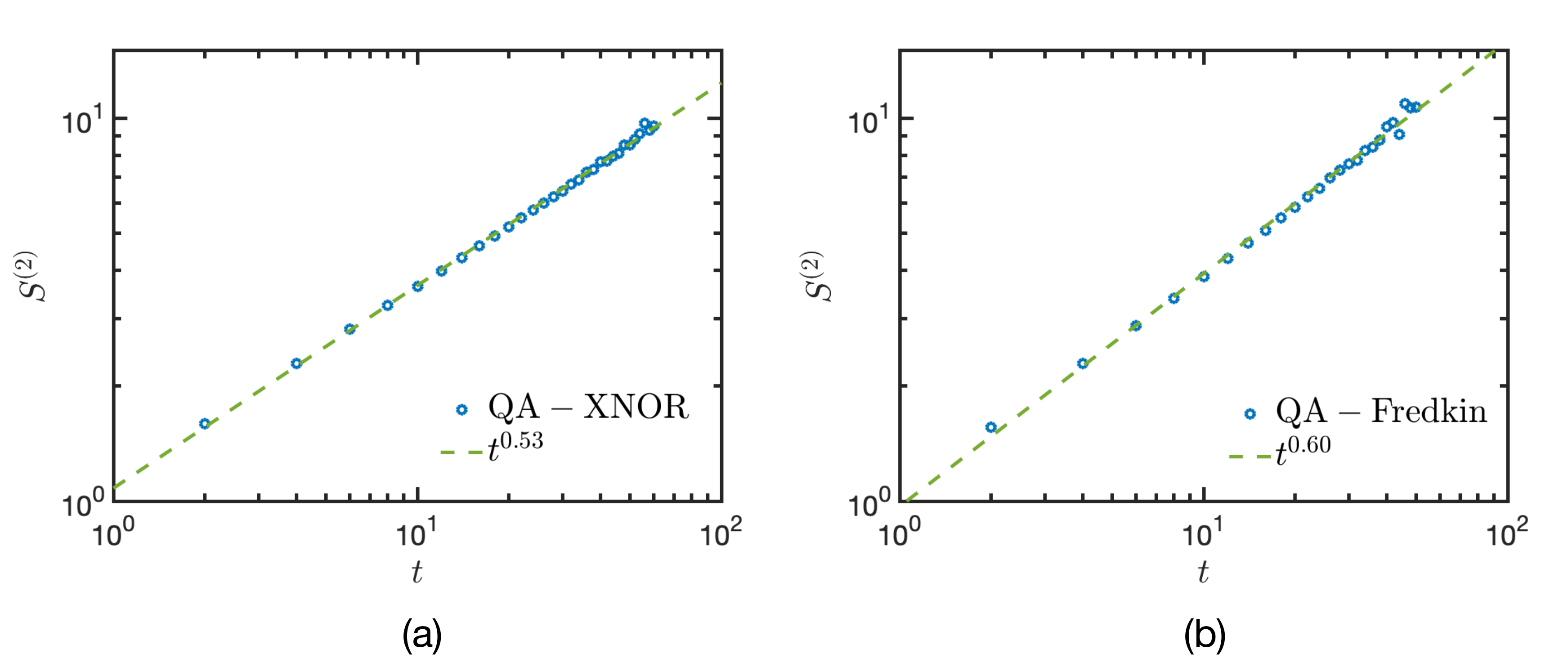}
\caption{Growth of $S^{(2)}$ for (a) the XNOR model and (b) the Fredkin model. Numerical results show that the dynamics of the second R\'enyi entropy is in good agreement with $S^{(2)} \propto t^{1/2}$ for the XNOR model, and $S^{(2)}\propto t^{0.6}$ for the Fredkin model, respectively. In both cases, the exponents are different from $1/z$, where $z$ is extracted from $C(x,t)$, as shown in Fig.~\ref{fig:correlation}. Numerical simulations are performed for systems with $L=50$ qubits and averaged over $10^4$ circuit realizations.}
\label{fig:renyi2} 
\end{figure}

We numerically compute the purity and the second R\'enyi entropy for both QA circuits by sampling from Eq.~(\ref{eq:purity}). As shown in Fig.~\ref{fig:renyi2}, the second R\'enyi entropy grows diffusively with $S^{(2)}\propto t^{1/2}$ for the XNOR model, and superdiffusively with $S^{(2)} \propto t^{0.6}$ for the Fredkin model. In both models, the dynamics of $S^{(2)}$ differ from spin transport as diagnosed from the spin correlators. To understand this distinction, it is useful to briefly review the argument that led to the diffusive growth of $S^{(2)}$ in systems with U(1) charge conservation~\cite{PhysRevLett.122.250602}. Under unitary evolution, there are contributions to $|\psi(t)\rangle$ coming from rare trajectories where no particle passes through the central bond up to time $t$. As particles move diffusively, this probability decays as $e^{-\gamma t^{1/2}}$, which further lower bounds the largest eigenvalue of the reduced density matrix $\rho_A$: $\Lambda_{\rm max} \geq e^{-\gamma t^{1/2}}$. Hence, the second R\'enyi entropy satisfies $S^{(2)} \leq 2S^{(\infty)} := -{\rm ln} \Lambda_{\rm max} \leq \mathcal{O}(t^{1/2})$ and can grow no faster than diffusively.

While the above argument generalizes to transport types other than diffusion, one immediately notices a key distinction from the spin correlators for the XNOR model. As we show previously, the spin correlator in the XNOR model is described by an emergent tracer dynamics of tagged particles, here the dynamics of higher R\'enyi entropies only care about collective transport. In the nearest-neighbor simple exclusion picture of the XNOR model, collective transport of particles is diffusive. 
The scenario here is also reminiscent of the distinction between spin and energy transport in the integrable easy-axis $XXZ$ model: while spin transport is diffusive due to screening of the quasiparticles (magnons), energy transport is still ballistic since quasiparticles do propagate ballistically~\cite{ljubotina2017spin, PhysRevLett.106.220601, PhysRevLett.122.127202}. This provides a simple argument for the distinct behaviors between transport and R\'enyi entropy dynamics for the XNOR model.

On the other hand, a theoretical prediction for the superdiffusive growth of $S^{(2)}$ in the Fredkin model still remains elusive due to a lack of understanding of the nature of this universality class. The observed distinction between transport and R\'enyi entropy dynamics seems to suggest that there might be a similar picture underlying the Fredkin model. However, notice a key difference from the XNOR model: the inverse of the exponent governing the R\'enyi entropy dynamics is not simply related to the dynamical exponent $z=\frac{8}{3}$ by a factor of two. This indicates that the picture behind the $z=\frac{8}{3}$ scaling is perhaps more complicated than a screening of some superdiffusive ``magnons" with $z'=\frac{4}{3}$.

In the Supplemental Material, we present additional numerical results on U(1)-symmetric QA circuits with other types of kinetic constraints. Models that we explore feature diffusive, localized, and quasilocalized dynamics. In these models, we instead find that the dynamics of the second R\'enyi entropy are consistent with those diagnosed using spin correlators and hence in agreement with Eq.~(\ref{eq:renyi_z}).

{\it Discussion.--} We provide concrete examples showing that the dynamics of higher R\'enyi entropies in U(1) symmetric kinetically constrained systems can be different from those diagnosed using correlation functions of the conserved charges. Since the emergent tracer dynamics picture for the XNOR model is also valid for Haar random circuits~\cite{feldmeier2022emergent}, we expect that this distinction remains for more generic unitary evolutions other than QA dynamics. While a theoretical understanding of the Fredkin dynamics is still unclear, the fact that the dynamics of $S^{(2)}$ and transport also show a distinction suggests that there might be a similar picture underlying the Fredkin dynamics, although important differences are noted. Our results thus may also shed light on future investigations of the Fredkin model. A more systematic approach for identifying models where relation~(\ref{eq:renyi_z}) does not hold is also an interesting direction for future work.

I would like to thank Michael Knap and Xiao Chen for helpful discussions. This work is supported by a startup fund at Peking University. The numerical calculations were performed on the Boston University Shared Computing Cluster, which is administered by Boston University Research Computing Services.
 

\bibliography{reference}

\begin{thebibliography}{38}%
\makeatletter
\providecommand \@ifxundefined [1]{%
 \@ifx{#1\undefined}
}%
\providecommand \@ifnum [1]{%
 \ifnum #1\expandafter \@firstoftwo
 \else \expandafter \@secondoftwo
 \fi
}%
\providecommand \@ifx [1]{%
 \ifx #1\expandafter \@firstoftwo
 \else \expandafter \@secondoftwo
 \fi
}%
\providecommand \natexlab [1]{#1}%
\providecommand \enquote  [1]{``#1''}%
\providecommand \bibnamefont  [1]{#1}%
\providecommand \bibfnamefont [1]{#1}%
\providecommand \citenamefont [1]{#1}%
\providecommand \href@noop [0]{\@secondoftwo}%
\providecommand \href [0]{\begingroup \@sanitize@url \@href}%
\providecommand \@href[1]{\@@startlink{#1}\@@href}%
\providecommand \@@href[1]{\endgroup#1\@@endlink}%
\providecommand \@sanitize@url [0]{\catcode `\\12\catcode `\$12\catcode
  `\&12\catcode `\#12\catcode `\^12\catcode `\_12\catcode `\%12\relax}%
\providecommand \@@startlink[1]{}%
\providecommand \@@endlink[0]{}%
\providecommand \url  [0]{\begingroup\@sanitize@url \@url }%
\providecommand \@url [1]{\endgroup\@href {#1}{\urlprefix }}%
\providecommand \urlprefix  [0]{URL }%
\providecommand \Eprint [0]{\href }%
\providecommand \doibase [0]{https://doi.org/}%
\providecommand \selectlanguage [0]{\@gobble}%
\providecommand \bibinfo  [0]{\@secondoftwo}%
\providecommand \bibfield  [0]{\@secondoftwo}%
\providecommand \translation [1]{[#1]}%
\providecommand \BibitemOpen [0]{}%
\providecommand \bibitemStop [0]{}%
\providecommand \bibitemNoStop [0]{.\EOS\space}%
\providecommand \EOS [0]{\spacefactor3000\relax}%
\providecommand \BibitemShut  [1]{\csname bibitem#1\endcsname}%
\let\auto@bib@innerbib\@empty
\bibitem [{\citenamefont {Rigol}\ \emph {et~al.}(2008)\citenamefont {Rigol},
  \citenamefont {Dunjko},\ and\ \citenamefont
  {Olshanii}}]{rigol2008thermalization}%
  \BibitemOpen
  \bibfield  {author} {\bibinfo {author} {\bibfnamefont {M.}~\bibnamefont
  {Rigol}}, \bibinfo {author} {\bibfnamefont {V.}~\bibnamefont {Dunjko}},\ and\
  \bibinfo {author} {\bibfnamefont {M.}~\bibnamefont {Olshanii}},\ }\href@noop
  {} {\bibfield  {journal} {\bibinfo  {journal} {Nature}\ }\textbf {\bibinfo
  {volume} {452}},\ \bibinfo {pages} {854} (\bibinfo {year}
  {2008})}\BibitemShut {NoStop}%
\bibitem [{\citenamefont {D'Alessio}\ \emph {et~al.}(2016)\citenamefont
  {D'Alessio}, \citenamefont {Kafri}, \citenamefont {Polkovnikov},\ and\
  \citenamefont {Rigol}}]{d2016quantum}%
  \BibitemOpen
  \bibfield  {author} {\bibinfo {author} {\bibfnamefont {L.}~\bibnamefont
  {D'Alessio}}, \bibinfo {author} {\bibfnamefont {Y.}~\bibnamefont {Kafri}},
  \bibinfo {author} {\bibfnamefont {A.}~\bibnamefont {Polkovnikov}},\ and\
  \bibinfo {author} {\bibfnamefont {M.}~\bibnamefont {Rigol}},\ }\href@noop {}
  {\bibfield  {journal} {\bibinfo  {journal} {Advances in Physics}\ }\textbf
  {\bibinfo {volume} {65}},\ \bibinfo {pages} {239} (\bibinfo {year}
  {2016})}\BibitemShut {NoStop}%
\bibitem [{\citenamefont {Gogolin}\ and\ \citenamefont
  {Eisert}(2016)}]{gogolin2016equilibration}%
  \BibitemOpen
  \bibfield  {author} {\bibinfo {author} {\bibfnamefont {C.}~\bibnamefont
  {Gogolin}}\ and\ \bibinfo {author} {\bibfnamefont {J.}~\bibnamefont
  {Eisert}},\ }\href@noop {} {\bibfield  {journal} {\bibinfo  {journal}
  {Reports on Progress in Physics}\ }\textbf {\bibinfo {volume} {79}},\
  \bibinfo {pages} {056001} (\bibinfo {year} {2016})}\BibitemShut {NoStop}%
\bibitem [{\citenamefont {Preskill}(2018)}]{preskill2018quantum}%
  \BibitemOpen
  \bibfield  {author} {\bibinfo {author} {\bibfnamefont {J.}~\bibnamefont
  {Preskill}},\ }\href@noop {} {\bibfield  {journal} {\bibinfo  {journal}
  {Quantum}\ }\textbf {\bibinfo {volume} {2}},\ \bibinfo {pages} {79} (\bibinfo
  {year} {2018})}\BibitemShut {NoStop}%
\bibitem [{\citenamefont {Mukerjee}\ \emph {et~al.}(2006)\citenamefont
  {Mukerjee}, \citenamefont {Oganesyan},\ and\ \citenamefont
  {Huse}}]{PhysRevB.73.035113}%
  \BibitemOpen
  \bibfield  {author} {\bibinfo {author} {\bibfnamefont {S.}~\bibnamefont
  {Mukerjee}}, \bibinfo {author} {\bibfnamefont {V.}~\bibnamefont
  {Oganesyan}},\ and\ \bibinfo {author} {\bibfnamefont {D.}~\bibnamefont
  {Huse}},\ }\href {https://doi.org/10.1103/PhysRevB.73.035113} {\bibfield
  {journal} {\bibinfo  {journal} {Phys. Rev. B}\ }\textbf {\bibinfo {volume}
  {73}},\ \bibinfo {pages} {035113} (\bibinfo {year} {2006})}\BibitemShut
  {NoStop}%
\bibitem [{\citenamefont {Lux}\ \emph {et~al.}(2014)\citenamefont {Lux},
  \citenamefont {M\"uller}, \citenamefont {Mitra},\ and\ \citenamefont
  {Rosch}}]{PhysRevA.89.053608}%
  \BibitemOpen
  \bibfield  {author} {\bibinfo {author} {\bibfnamefont {J.}~\bibnamefont
  {Lux}}, \bibinfo {author} {\bibfnamefont {J.}~\bibnamefont {M\"uller}},
  \bibinfo {author} {\bibfnamefont {A.}~\bibnamefont {Mitra}},\ and\ \bibinfo
  {author} {\bibfnamefont {A.}~\bibnamefont {Rosch}},\ }\href
  {https://doi.org/10.1103/PhysRevA.89.053608} {\bibfield  {journal} {\bibinfo
  {journal} {Phys. Rev. A}\ }\textbf {\bibinfo {volume} {89}},\ \bibinfo
  {pages} {053608} (\bibinfo {year} {2014})}\BibitemShut {NoStop}%
\bibitem [{\citenamefont {Khemani}\ \emph {et~al.}(2018)\citenamefont
  {Khemani}, \citenamefont {Vishwanath},\ and\ \citenamefont
  {Huse}}]{PhysRevX.8.031057}%
  \BibitemOpen
  \bibfield  {author} {\bibinfo {author} {\bibfnamefont {V.}~\bibnamefont
  {Khemani}}, \bibinfo {author} {\bibfnamefont {A.}~\bibnamefont
  {Vishwanath}},\ and\ \bibinfo {author} {\bibfnamefont {D.~A.}\ \bibnamefont
  {Huse}},\ }\href {https://doi.org/10.1103/PhysRevX.8.031057} {\bibfield
  {journal} {\bibinfo  {journal} {Phys. Rev. X}\ }\textbf {\bibinfo {volume}
  {8}},\ \bibinfo {pages} {031057} (\bibinfo {year} {2018})}\BibitemShut
  {NoStop}%
\bibitem [{\citenamefont {Rakovszky}\ \emph {et~al.}(2018)\citenamefont
  {Rakovszky}, \citenamefont {Pollmann},\ and\ \citenamefont {von
  Keyserlingk}}]{PhysRevX.8.031058}%
  \BibitemOpen
  \bibfield  {author} {\bibinfo {author} {\bibfnamefont {T.}~\bibnamefont
  {Rakovszky}}, \bibinfo {author} {\bibfnamefont {F.}~\bibnamefont
  {Pollmann}},\ and\ \bibinfo {author} {\bibfnamefont {C.~W.}\ \bibnamefont
  {von Keyserlingk}},\ }\href {https://doi.org/10.1103/PhysRevX.8.031058}
  {\bibfield  {journal} {\bibinfo  {journal} {Phys. Rev. X}\ }\textbf {\bibinfo
  {volume} {8}},\ \bibinfo {pages} {031058} (\bibinfo {year}
  {2018})}\BibitemShut {NoStop}%
\bibitem [{\citenamefont {Gromov}\ \emph {et~al.}(2020)\citenamefont {Gromov},
  \citenamefont {Lucas},\ and\ \citenamefont
  {Nandkishore}}]{PhysRevResearch.2.033124}%
  \BibitemOpen
  \bibfield  {author} {\bibinfo {author} {\bibfnamefont {A.}~\bibnamefont
  {Gromov}}, \bibinfo {author} {\bibfnamefont {A.}~\bibnamefont {Lucas}},\ and\
  \bibinfo {author} {\bibfnamefont {R.~M.}\ \bibnamefont {Nandkishore}},\
  }\href {https://doi.org/10.1103/PhysRevResearch.2.033124} {\bibfield
  {journal} {\bibinfo  {journal} {Phys. Rev. Research}\ }\textbf {\bibinfo
  {volume} {2}},\ \bibinfo {pages} {033124} (\bibinfo {year}
  {2020})}\BibitemShut {NoStop}%
\bibitem [{\citenamefont {Glorioso}\ \emph {et~al.}(2022)\citenamefont
  {Glorioso}, \citenamefont {Guo}, \citenamefont {Rodriguez-Nieva},\ and\
  \citenamefont {Lucas}}]{glorioso2022breakdown}%
  \BibitemOpen
  \bibfield  {author} {\bibinfo {author} {\bibfnamefont {P.}~\bibnamefont
  {Glorioso}}, \bibinfo {author} {\bibfnamefont {J.}~\bibnamefont {Guo}},
  \bibinfo {author} {\bibfnamefont {J.~F.}\ \bibnamefont {Rodriguez-Nieva}},\
  and\ \bibinfo {author} {\bibfnamefont {A.}~\bibnamefont {Lucas}},\
  }\href@noop {} {\bibfield  {journal} {\bibinfo  {journal} {Nature Physics}\
  ,\ \bibinfo {pages} {1}} (\bibinfo {year} {2022})}\BibitemShut {NoStop}%
\bibitem [{\citenamefont {Ilievski}\ \emph {et~al.}(2018)\citenamefont
  {Ilievski}, \citenamefont {De~Nardis}, \citenamefont {Medenjak},\ and\
  \citenamefont {Prosen}}]{PhysRevLett.121.230602}%
  \BibitemOpen
  \bibfield  {author} {\bibinfo {author} {\bibfnamefont {E.}~\bibnamefont
  {Ilievski}}, \bibinfo {author} {\bibfnamefont {J.}~\bibnamefont {De~Nardis}},
  \bibinfo {author} {\bibfnamefont {M.}~\bibnamefont {Medenjak}},\ and\
  \bibinfo {author} {\bibfnamefont {T.~c.~v.}\ \bibnamefont {Prosen}},\ }\href
  {https://doi.org/10.1103/PhysRevLett.121.230602} {\bibfield  {journal}
  {\bibinfo  {journal} {Phys. Rev. Lett.}\ }\textbf {\bibinfo {volume} {121}},\
  \bibinfo {pages} {230602} (\bibinfo {year} {2018})}\BibitemShut {NoStop}%
\bibitem [{\citenamefont {Ljubotina}\ \emph {et~al.}(2017)\citenamefont
  {Ljubotina}, \citenamefont {{\v{Z}}nidari{\v{c}}},\ and\ \citenamefont
  {Prosen}}]{ljubotina2017spin}%
  \BibitemOpen
  \bibfield  {author} {\bibinfo {author} {\bibfnamefont {M.}~\bibnamefont
  {Ljubotina}}, \bibinfo {author} {\bibfnamefont {M.}~\bibnamefont
  {{\v{Z}}nidari{\v{c}}}},\ and\ \bibinfo {author} {\bibfnamefont
  {T.}~\bibnamefont {Prosen}},\ }\href@noop {} {\bibfield  {journal} {\bibinfo
  {journal} {Nature communications}\ }\textbf {\bibinfo {volume} {8}},\
  \bibinfo {pages} {1} (\bibinfo {year} {2017})}\BibitemShut {NoStop}%
\bibitem [{\citenamefont {Gopalakrishnan}\ \emph {et~al.}(2019)\citenamefont
  {Gopalakrishnan}, \citenamefont {Vasseur},\ and\ \citenamefont
  {Ware}}]{gopalakrishnan2019anomalous}%
  \BibitemOpen
  \bibfield  {author} {\bibinfo {author} {\bibfnamefont {S.}~\bibnamefont
  {Gopalakrishnan}}, \bibinfo {author} {\bibfnamefont {R.}~\bibnamefont
  {Vasseur}},\ and\ \bibinfo {author} {\bibfnamefont {B.}~\bibnamefont
  {Ware}},\ }\href@noop {} {\bibfield  {journal} {\bibinfo  {journal}
  {Proceedings of the National Academy of Sciences}\ }\textbf {\bibinfo
  {volume} {116}},\ \bibinfo {pages} {16250} (\bibinfo {year}
  {2019})}\BibitemShut {NoStop}%
\bibitem [{\citenamefont {De~Nardis}\ \emph
  {et~al.}(2021{\natexlab{a}})\citenamefont {De~Nardis}, \citenamefont
  {Gopalakrishnan}, \citenamefont {Vasseur},\ and\ \citenamefont
  {Ware}}]{PhysRevLett.127.057201}%
  \BibitemOpen
  \bibfield  {author} {\bibinfo {author} {\bibfnamefont {J.}~\bibnamefont
  {De~Nardis}}, \bibinfo {author} {\bibfnamefont {S.}~\bibnamefont
  {Gopalakrishnan}}, \bibinfo {author} {\bibfnamefont {R.}~\bibnamefont
  {Vasseur}},\ and\ \bibinfo {author} {\bibfnamefont {B.}~\bibnamefont
  {Ware}},\ }\href {https://doi.org/10.1103/PhysRevLett.127.057201} {\bibfield
  {journal} {\bibinfo  {journal} {Phys. Rev. Lett.}\ }\textbf {\bibinfo
  {volume} {127}},\ \bibinfo {pages} {057201} (\bibinfo {year}
  {2021}{\natexlab{a}})}\BibitemShut {NoStop}%
\bibitem [{\citenamefont {Ilievski}\ \emph {et~al.}(2021)\citenamefont
  {Ilievski}, \citenamefont {De~Nardis}, \citenamefont {Gopalakrishnan},
  \citenamefont {Vasseur},\ and\ \citenamefont {Ware}}]{PhysRevX.11.031023}%
  \BibitemOpen
  \bibfield  {author} {\bibinfo {author} {\bibfnamefont {E.}~\bibnamefont
  {Ilievski}}, \bibinfo {author} {\bibfnamefont {J.}~\bibnamefont {De~Nardis}},
  \bibinfo {author} {\bibfnamefont {S.}~\bibnamefont {Gopalakrishnan}},
  \bibinfo {author} {\bibfnamefont {R.}~\bibnamefont {Vasseur}},\ and\ \bibinfo
  {author} {\bibfnamefont {B.}~\bibnamefont {Ware}},\ }\href
  {https://doi.org/10.1103/PhysRevX.11.031023} {\bibfield  {journal} {\bibinfo
  {journal} {Phys. Rev. X}\ }\textbf {\bibinfo {volume} {11}},\ \bibinfo
  {pages} {031023} (\bibinfo {year} {2021})}\BibitemShut {NoStop}%
\bibitem [{\citenamefont {Gopalakrishnan}\ and\ \citenamefont
  {Vasseur}(2019)}]{PhysRevLett.122.127202}%
  \BibitemOpen
  \bibfield  {author} {\bibinfo {author} {\bibfnamefont {S.}~\bibnamefont
  {Gopalakrishnan}}\ and\ \bibinfo {author} {\bibfnamefont {R.}~\bibnamefont
  {Vasseur}},\ }\href {https://doi.org/10.1103/PhysRevLett.122.127202}
  {\bibfield  {journal} {\bibinfo  {journal} {Phys. Rev. Lett.}\ }\textbf
  {\bibinfo {volume} {122}},\ \bibinfo {pages} {127202} (\bibinfo {year}
  {2019})}\BibitemShut {NoStop}%
\bibitem [{\citenamefont {De~Nardis}\ \emph
  {et~al.}(2021{\natexlab{b}})\citenamefont {De~Nardis}, \citenamefont
  {Gopalakrishnan}, \citenamefont {Vasseur},\ and\ \citenamefont
  {Ware}}]{de2021subdiffusive}%
  \BibitemOpen
  \bibfield  {author} {\bibinfo {author} {\bibfnamefont {J.}~\bibnamefont
  {De~Nardis}}, \bibinfo {author} {\bibfnamefont {S.}~\bibnamefont
  {Gopalakrishnan}}, \bibinfo {author} {\bibfnamefont {R.}~\bibnamefont
  {Vasseur}},\ and\ \bibinfo {author} {\bibfnamefont {B.}~\bibnamefont
  {Ware}},\ }\href@noop {} {\bibfield  {journal} {\bibinfo  {journal} {arXiv
  preprint arXiv:2109.13251}\ } (\bibinfo {year}
  {2021}{\natexlab{b}})}\BibitemShut {NoStop}%
\bibitem [{\citenamefont {Kim}\ and\ \citenamefont
  {Huse}(2013)}]{PhysRevLett.111.127205}%
  \BibitemOpen
  \bibfield  {author} {\bibinfo {author} {\bibfnamefont {H.}~\bibnamefont
  {Kim}}\ and\ \bibinfo {author} {\bibfnamefont {D.~A.}\ \bibnamefont {Huse}},\
  }\href {https://doi.org/10.1103/PhysRevLett.111.127205} {\bibfield  {journal}
  {\bibinfo  {journal} {Phys. Rev. Lett.}\ }\textbf {\bibinfo {volume} {111}},\
  \bibinfo {pages} {127205} (\bibinfo {year} {2013})}\BibitemShut {NoStop}%
\bibitem [{\citenamefont {Rakovszky}\ \emph {et~al.}(2019)\citenamefont
  {Rakovszky}, \citenamefont {Pollmann},\ and\ \citenamefont {von
  Keyserlingk}}]{PhysRevLett.122.250602}%
  \BibitemOpen
  \bibfield  {author} {\bibinfo {author} {\bibfnamefont {T.}~\bibnamefont
  {Rakovszky}}, \bibinfo {author} {\bibfnamefont {F.}~\bibnamefont
  {Pollmann}},\ and\ \bibinfo {author} {\bibfnamefont {C.~W.}\ \bibnamefont
  {von Keyserlingk}},\ }\href {https://doi.org/10.1103/PhysRevLett.122.250602}
  {\bibfield  {journal} {\bibinfo  {journal} {Phys. Rev. Lett.}\ }\textbf
  {\bibinfo {volume} {122}},\ \bibinfo {pages} {250602} (\bibinfo {year}
  {2019})}\BibitemShut {NoStop}%
\bibitem [{\citenamefont
  {{\v{Z}}nidari{\v{c}}}(2020)}]{vznidarivc2020entanglement}%
  \BibitemOpen
  \bibfield  {author} {\bibinfo {author} {\bibfnamefont {M.}~\bibnamefont
  {{\v{Z}}nidari{\v{c}}}},\ }\href@noop {} {\bibfield  {journal} {\bibinfo
  {journal} {Communications Physics}\ }\textbf {\bibinfo {volume} {3}},\
  \bibinfo {pages} {1} (\bibinfo {year} {2020})}\BibitemShut {NoStop}%
\bibitem [{\citenamefont {Huang}(2020)}]{huang2020dynamics}%
  \BibitemOpen
  \bibfield  {author} {\bibinfo {author} {\bibfnamefont {Y.}~\bibnamefont
  {Huang}},\ }\href@noop {} {\bibfield  {journal} {\bibinfo  {journal} {IOP
  SciNotes}\ }\textbf {\bibinfo {volume} {1}},\ \bibinfo {pages} {035205}
  (\bibinfo {year} {2020})}\BibitemShut {NoStop}%
\bibitem [{\citenamefont {Zhou}\ and\ \citenamefont
  {Ludwig}(2020)}]{PhysRevResearch.2.033020}%
  \BibitemOpen
  \bibfield  {author} {\bibinfo {author} {\bibfnamefont {T.}~\bibnamefont
  {Zhou}}\ and\ \bibinfo {author} {\bibfnamefont {A.~W.~W.}\ \bibnamefont
  {Ludwig}},\ }\href {https://doi.org/10.1103/PhysRevResearch.2.033020}
  {\bibfield  {journal} {\bibinfo  {journal} {Phys. Rev. Research}\ }\textbf
  {\bibinfo {volume} {2}},\ \bibinfo {pages} {033020} (\bibinfo {year}
  {2020})}\BibitemShut {NoStop}%
\bibitem [{\citenamefont {Richter}\ \emph {et~al.}(2022)\citenamefont
  {Richter}, \citenamefont {Lunt},\ and\ \citenamefont
  {Pal}}]{richter2022transport}%
  \BibitemOpen
  \bibfield  {author} {\bibinfo {author} {\bibfnamefont {J.}~\bibnamefont
  {Richter}}, \bibinfo {author} {\bibfnamefont {O.}~\bibnamefont {Lunt}},\ and\
  \bibinfo {author} {\bibfnamefont {A.}~\bibnamefont {Pal}},\ }\href@noop {}
  {\bibfield  {journal} {\bibinfo  {journal} {arXiv preprint arXiv:2205.06309}\
  } (\bibinfo {year} {2022})}\BibitemShut {NoStop}%
\bibitem [{\citenamefont {Singh}\ \emph {et~al.}(2021)\citenamefont {Singh},
  \citenamefont {Ware}, \citenamefont {Vasseur},\ and\ \citenamefont
  {Friedman}}]{PhysRevLett.127.230602}%
  \BibitemOpen
  \bibfield  {author} {\bibinfo {author} {\bibfnamefont {H.}~\bibnamefont
  {Singh}}, \bibinfo {author} {\bibfnamefont {B.~A.}\ \bibnamefont {Ware}},
  \bibinfo {author} {\bibfnamefont {R.}~\bibnamefont {Vasseur}},\ and\ \bibinfo
  {author} {\bibfnamefont {A.~J.}\ \bibnamefont {Friedman}},\ }\href
  {https://doi.org/10.1103/PhysRevLett.127.230602} {\bibfield  {journal}
  {\bibinfo  {journal} {Phys. Rev. Lett.}\ }\textbf {\bibinfo {volume} {127}},\
  \bibinfo {pages} {230602} (\bibinfo {year} {2021})}\BibitemShut {NoStop}%
\bibitem [{\citenamefont {Yang}\ \emph {et~al.}(2020)\citenamefont {Yang},
  \citenamefont {Liu}, \citenamefont {Gorshkov},\ and\ \citenamefont
  {Iadecola}}]{PhysRevLett.124.207602}%
  \BibitemOpen
  \bibfield  {author} {\bibinfo {author} {\bibfnamefont {Z.-C.}\ \bibnamefont
  {Yang}}, \bibinfo {author} {\bibfnamefont {F.}~\bibnamefont {Liu}}, \bibinfo
  {author} {\bibfnamefont {A.~V.}\ \bibnamefont {Gorshkov}},\ and\ \bibinfo
  {author} {\bibfnamefont {T.}~\bibnamefont {Iadecola}},\ }\href
  {https://doi.org/10.1103/PhysRevLett.124.207602} {\bibfield  {journal}
  {\bibinfo  {journal} {Phys. Rev. Lett.}\ }\textbf {\bibinfo {volume} {124}},\
  \bibinfo {pages} {207602} (\bibinfo {year} {2020})}\BibitemShut {NoStop}%
\bibitem [{\citenamefont {Langlett}\ and\ \citenamefont
  {Xu}(2021)}]{PhysRevB.103.L220304}%
  \BibitemOpen
  \bibfield  {author} {\bibinfo {author} {\bibfnamefont {C.~M.}\ \bibnamefont
  {Langlett}}\ and\ \bibinfo {author} {\bibfnamefont {S.}~\bibnamefont {Xu}},\
  }\href {https://doi.org/10.1103/PhysRevB.103.L220304} {\bibfield  {journal}
  {\bibinfo  {journal} {Phys. Rev. B}\ }\textbf {\bibinfo {volume} {103}},\
  \bibinfo {pages} {L220304} (\bibinfo {year} {2021})}\BibitemShut {NoStop}%
\bibitem [{\citenamefont {Feldmeier}\ \emph {et~al.}(2022)\citenamefont
  {Feldmeier}, \citenamefont {Witczak-Krempa},\ and\ \citenamefont
  {Knap}}]{feldmeier2022emergent}%
  \BibitemOpen
  \bibfield  {author} {\bibinfo {author} {\bibfnamefont {J.}~\bibnamefont
  {Feldmeier}}, \bibinfo {author} {\bibfnamefont {W.}~\bibnamefont
  {Witczak-Krempa}},\ and\ \bibinfo {author} {\bibfnamefont {M.}~\bibnamefont
  {Knap}},\ }\href@noop {} {\bibfield  {journal} {\bibinfo  {journal} {arXiv
  preprint arXiv:2205.07901}\ } (\bibinfo {year} {2022})}\BibitemShut {NoStop}%
\bibitem [{\citenamefont {\ifmmode \check{Z}\else
  \v{Z}\fi{}nidari\ifmmode~\check{c}\else
  \v{c}\fi{}}(2011)}]{PhysRevLett.106.220601}%
  \BibitemOpen
  \bibfield  {author} {\bibinfo {author} {\bibfnamefont {M.}~\bibnamefont
  {\ifmmode \check{Z}\else \v{Z}\fi{}nidari\ifmmode~\check{c}\else
  \v{c}\fi{}}},\ }\href {https://doi.org/10.1103/PhysRevLett.106.220601}
  {\bibfield  {journal} {\bibinfo  {journal} {Phys. Rev. Lett.}\ }\textbf
  {\bibinfo {volume} {106}},\ \bibinfo {pages} {220601} (\bibinfo {year}
  {2011})}\BibitemShut {NoStop}%
\bibitem [{\citenamefont {Iaconis}\ \emph {et~al.}(2019)\citenamefont
  {Iaconis}, \citenamefont {Vijay},\ and\ \citenamefont
  {Nandkishore}}]{PhysRevB.100.214301}%
  \BibitemOpen
  \bibfield  {author} {\bibinfo {author} {\bibfnamefont {J.}~\bibnamefont
  {Iaconis}}, \bibinfo {author} {\bibfnamefont {S.}~\bibnamefont {Vijay}},\
  and\ \bibinfo {author} {\bibfnamefont {R.}~\bibnamefont {Nandkishore}},\
  }\href {https://doi.org/10.1103/PhysRevB.100.214301} {\bibfield  {journal}
  {\bibinfo  {journal} {Phys. Rev. B}\ }\textbf {\bibinfo {volume} {100}},\
  \bibinfo {pages} {214301} (\bibinfo {year} {2019})}\BibitemShut {NoStop}%
\bibitem [{\citenamefont {Iaconis}(2021)}]{PRXQuantum.2.010329}%
  \BibitemOpen
  \bibfield  {author} {\bibinfo {author} {\bibfnamefont {J.}~\bibnamefont
  {Iaconis}},\ }\href {https://doi.org/10.1103/PRXQuantum.2.010329} {\bibfield
  {journal} {\bibinfo  {journal} {PRX Quantum}\ }\textbf {\bibinfo {volume}
  {2}},\ \bibinfo {pages} {010329} (\bibinfo {year} {2021})}\BibitemShut
  {NoStop}%
\bibitem [{\citenamefont {Iaconis}\ \emph {et~al.}(2020)\citenamefont
  {Iaconis}, \citenamefont {Lucas},\ and\ \citenamefont
  {Chen}}]{PhysRevB.102.224311}%
  \BibitemOpen
  \bibfield  {author} {\bibinfo {author} {\bibfnamefont {J.}~\bibnamefont
  {Iaconis}}, \bibinfo {author} {\bibfnamefont {A.}~\bibnamefont {Lucas}},\
  and\ \bibinfo {author} {\bibfnamefont {X.}~\bibnamefont {Chen}},\ }\href
  {https://doi.org/10.1103/PhysRevB.102.224311} {\bibfield  {journal} {\bibinfo
   {journal} {Phys. Rev. B}\ }\textbf {\bibinfo {volume} {102}},\ \bibinfo
  {pages} {224311} (\bibinfo {year} {2020})}\BibitemShut {NoStop}%
\bibitem [{\citenamefont {Gopalakrishnan}\ and\ \citenamefont
  {Zakirov}(2018)}]{gopalakrishnan2018facilitated}%
  \BibitemOpen
  \bibfield  {author} {\bibinfo {author} {\bibfnamefont {S.}~\bibnamefont
  {Gopalakrishnan}}\ and\ \bibinfo {author} {\bibfnamefont {B.}~\bibnamefont
  {Zakirov}},\ }\href@noop {} {\bibfield  {journal} {\bibinfo  {journal}
  {Quantum Science and Technology}\ }\textbf {\bibinfo {volume} {3}},\ \bibinfo
  {pages} {044004} (\bibinfo {year} {2018})}\BibitemShut {NoStop}%
\bibitem [{\citenamefont {Han}\ and\ \citenamefont
  {Chen}(2022{\natexlab{a}})}]{PhysRevB.105.064306}%
  \BibitemOpen
  \bibfield  {author} {\bibinfo {author} {\bibfnamefont {Y.}~\bibnamefont
  {Han}}\ and\ \bibinfo {author} {\bibfnamefont {X.}~\bibnamefont {Chen}},\
  }\href {https://doi.org/10.1103/PhysRevB.105.064306} {\bibfield  {journal}
  {\bibinfo  {journal} {Phys. Rev. B}\ }\textbf {\bibinfo {volume} {105}},\
  \bibinfo {pages} {064306} (\bibinfo {year} {2022}{\natexlab{a}})}\BibitemShut
  {NoStop}%
\bibitem [{\citenamefont {Han}\ and\ \citenamefont
  {Chen}(2022{\natexlab{b}})}]{han2022entanglement}%
  \BibitemOpen
  \bibfield  {author} {\bibinfo {author} {\bibfnamefont {Y.}~\bibnamefont
  {Han}}\ and\ \bibinfo {author} {\bibfnamefont {X.}~\bibnamefont {Chen}},\
  }\href@noop {} {\bibfield  {journal} {\bibinfo  {journal} {arXiv preprint
  arXiv:2207.02165}\ } (\bibinfo {year} {2022}{\natexlab{b}})}\BibitemShut
  {NoStop}%
\bibitem [{\citenamefont {Chen}\ \emph {et~al.}(2017)\citenamefont {Chen},
  \citenamefont {Fradkin},\ and\ \citenamefont
  {Witczak-Krempa}}]{chen2017gapless}%
  \BibitemOpen
  \bibfield  {author} {\bibinfo {author} {\bibfnamefont {X.}~\bibnamefont
  {Chen}}, \bibinfo {author} {\bibfnamefont {E.}~\bibnamefont {Fradkin}},\ and\
  \bibinfo {author} {\bibfnamefont {W.}~\bibnamefont {Witczak-Krempa}},\
  }\href@noop {} {\bibfield  {journal} {\bibinfo  {journal} {Journal of Physics
  A: Mathematical and Theoretical}\ }\textbf {\bibinfo {volume} {50}},\
  \bibinfo {pages} {464002} (\bibinfo {year} {2017})}\BibitemShut {NoStop}%
\bibitem [{\citenamefont {Rakovszky}\ \emph {et~al.}(2020)\citenamefont
  {Rakovszky}, \citenamefont {Sala}, \citenamefont {Verresen}, \citenamefont
  {Knap},\ and\ \citenamefont {Pollmann}}]{PhysRevB.101.125126}%
  \BibitemOpen
  \bibfield  {author} {\bibinfo {author} {\bibfnamefont {T.}~\bibnamefont
  {Rakovszky}}, \bibinfo {author} {\bibfnamefont {P.}~\bibnamefont {Sala}},
  \bibinfo {author} {\bibfnamefont {R.}~\bibnamefont {Verresen}}, \bibinfo
  {author} {\bibfnamefont {M.}~\bibnamefont {Knap}},\ and\ \bibinfo {author}
  {\bibfnamefont {F.}~\bibnamefont {Pollmann}},\ }\href
  {https://doi.org/10.1103/PhysRevB.101.125126} {\bibfield  {journal} {\bibinfo
   {journal} {Phys. Rev. B}\ }\textbf {\bibinfo {volume} {101}},\ \bibinfo
  {pages} {125126} (\bibinfo {year} {2020})}\BibitemShut {NoStop}%
\bibitem [{\citenamefont {Alexander}\ and\ \citenamefont
  {Pincus}(1978)}]{PhysRevB.18.2011}%
  \BibitemOpen
  \bibfield  {author} {\bibinfo {author} {\bibfnamefont {S.}~\bibnamefont
  {Alexander}}\ and\ \bibinfo {author} {\bibfnamefont {P.}~\bibnamefont
  {Pincus}},\ }\href {https://doi.org/10.1103/PhysRevB.18.2011} {\bibfield
  {journal} {\bibinfo  {journal} {Phys. Rev. B}\ }\textbf {\bibinfo {volume}
  {18}},\ \bibinfo {pages} {2011} (\bibinfo {year} {1978})}\BibitemShut
  {NoStop}%
\bibitem [{\citenamefont {van Beijeren}\ \emph {et~al.}(1983)\citenamefont {van
  Beijeren}, \citenamefont {Kehr},\ and\ \citenamefont
  {Kutner}}]{PhysRevB.28.5711}%
  \BibitemOpen
  \bibfield  {author} {\bibinfo {author} {\bibfnamefont {H.}~\bibnamefont {van
  Beijeren}}, \bibinfo {author} {\bibfnamefont {K.~W.}\ \bibnamefont {Kehr}},\
  and\ \bibinfo {author} {\bibfnamefont {R.}~\bibnamefont {Kutner}},\ }\href
  {https://doi.org/10.1103/PhysRevB.28.5711} {\bibfield  {journal} {\bibinfo
  {journal} {Phys. Rev. B}\ }\textbf {\bibinfo {volume} {28}},\ \bibinfo
  {pages} {5711} (\bibinfo {year} {1983})}\BibitemShut {NoStop}%
\end{thebibliography}%


\newpage
\onecolumngrid
\appendix

\subsection*{Supplemental Material for ``Distinction Between Transport and R\'enyi Entropy Growth in Kinetically Constrained Models"}

\subsection{Additional results for other kinetically constrained models}

\begin{figure}[!h]
\includegraphics[width=.5\textwidth]{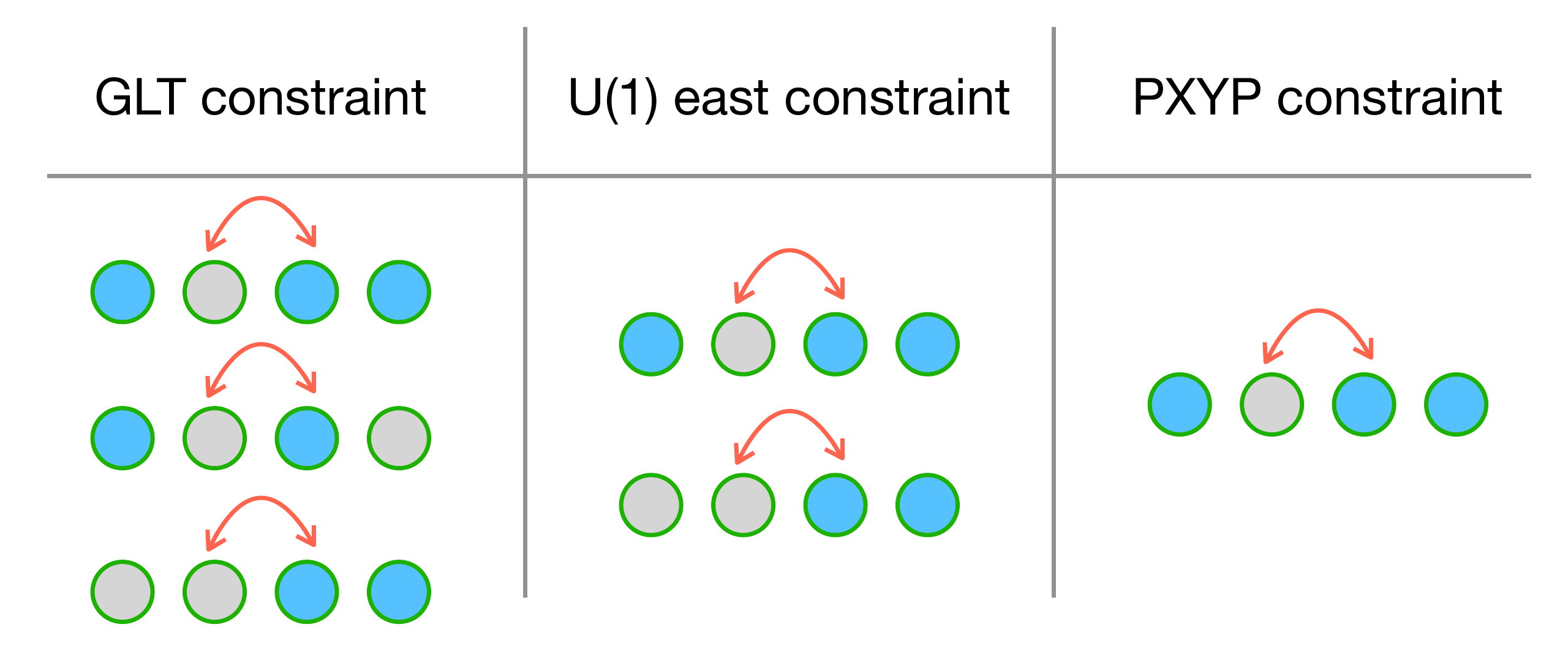}
\caption{Summary of allowed dynamical moves in other types of kinetically constrained models considered here.}
\label{fig:other} 
\end{figure}

We present additional numerical results on U(1)-symmetric QA circuits with other types of kinetic constraints. The models we consider are summarized in Fig.~\ref{fig:other}, in parallel with Ref.~\cite{PhysRevLett.127.230602}. The Gonçalves-Landim-Toninelli (GLT) model allows hopping between sites $i$ and $i+1$ only if either of their further-neighboring sites is occupied. The U(1) symmetric version of the East model allows hopping only if the further-neighboring site to the east (right) is occupied. Finally, the PXYP model allows hopping only if both further-neighboring sites are occupied, analogous to the PXP model. It was shown numerically in Ref.~\cite{PhysRevLett.127.230602} that these models exhibit distinct transport properties. The GLT model shows diffusive transport with $z=2$; the PXYP model is localized, and the U(1) East model has $z(t)$ growing without saturation, indicating quasilocalization.

\begin{figure}[!h]
\includegraphics[width=.9\textwidth]{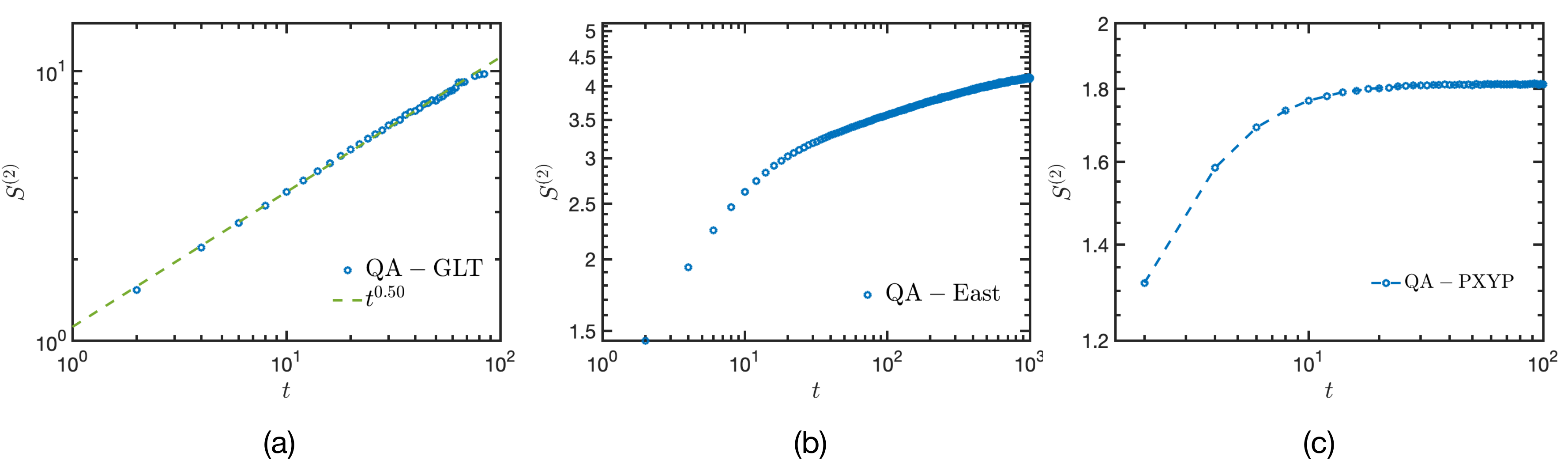}
\caption{Growth of $S^{(2)}$ for (a) the GLT model, (b) the U(1) East model, and (c) the PXYP model. In all three cases, the dynamics of $S^{(2)}$ are consistent with $S^{(2)}\propto t^{1/z}$. In the GLT model, $z=2$ (diffusion); in the U(1) East model, $z(t)$ increases without saturation (quasilocalization); in the PXYP model, $S^{(2)}$ itself saturates to an order one value (localization).}
\label{fig:renyi_other} 
\end{figure}

In Fig.~\ref{fig:renyi_other}, we show the dynamics of the second R\'enyi entropy for models listed above. We find that the results are all consistent with $S^{(2)} \propto t^{1/z}$. In particular, $S^{(2)}$ in the PXYP model quickly saturates to a value of order one, consistent with localization. $S^{(2)}$ in the GLT model grows as $t^{1/2}$, consistent with diffusion. Finally, $S^{(2)}$ in the U(1) East model grows with an exponent that decreases with time, suggesting $z(t)$ itself increases without saturation. Therefore, in contrast to the XNOR and Fredkin model studied in the main text, R\'enyi entropy dynamics in the three kinetically constrained models shown here exhibit behaviors consistent with spin transport.

\end{document}